\documentclass[pre,aps,twocolumn,superscriptaddress,floatfix,noshowpacs,nofootinbib,10pt]{revtex4-2}

\usepackage[utf8]{inputenc}     	
\usepackage{amsmath, amssymb, mathtools, mathrsfs}
\usepackage[colorlinks=true]{hyperref}
\hypersetup{colorlinks,linkcolor={red},citecolor={blue},urlcolor={blue}}  
\usepackage{color}
\usepackage{graphicx}
\usepackage[normalem]{ulem}
\usepackage{mathtools}
\usepackage{stmaryrd}
\usepackage{float}
\usepackage{textalpha}

\newcommand{\bs}[1]{{\boldsymbol{#1}}}

\newcommand{\br}{\bs{r}}

\newcommand{\bq}{\bs{q}}

\begin{document}
\title{Coarsening of binary Bose superfluids: an effective theory}
\author{Elisabeth Gliott}
\affiliation{Laboratoire Kastler Brossel, Sorbonne Universit\'{e}, CNRS, ENS-PSL Research University, 
Coll\`{e}ge de France; 4 Place Jussieu, 75005 Paris, France}

\author{Clara Piekarski}
\affiliation{Laboratoire Kastler Brossel, Sorbonne Universit\'{e}, CNRS, ENS-PSL Research University, 
Coll\`{e}ge de France; 4 Place Jussieu, 75005 Paris, France}

\author{Nicolas Cherroret}
\email{nicolas.cherroret@lkb.upmc.fr}
\affiliation{Laboratoire Kastler Brossel, Sorbonne Universit\'{e}, CNRS, ENS-PSL Research University, 
Coll\`{e}ge de France; 4 Place Jussieu, 75005 Paris, France}

\begin{abstract}
We derive an effective equation of motion for binary Bose mixtures, which generalizes the  Cahn-Hilliard description of classical binary fluids to superfluid systems. Within this approach,  based on a microscopic Hamiltonian formulation, we show that the domain growth law $L(t)\sim t^{2/3}$ observed in superfluid mixtures is not driven by hydrodynamic flows, but arises from the competition between interactions and quantum pressure. The effective theory allows us to derive key properties of superfluid coarsening, including domain growth and Porod's laws. This provides a new theoretical framework for understanding phase separation in superfluid mixtures.
\end{abstract}
\maketitle

\section{Introduction}

When a homogeneous system is quenched across a symmetry-breaking phase transition, its dynamics typically leads to the formation of growing  ordered-phase domains  \cite{Hohenberg1977, Cugliandolo2015}. At late times, this coarsening process is characterized by a self-similar dynamic scaling of correlations, governed by a single scale $L(t)\sim t^{1/z}$, where $z$  is a dynamic exponent. Coarsening has been studied  in various systems, such as the Ising \cite{Krapivsky2010}, XY \cite{Berthier2001} or $O(N)$ \cite{Chandran2013} models.  However, its theoretical description often relies on phenomenological approaches. Among these, the so-called models A and B of statistical physics have proven particularly useful \cite{Hohenberg1977, Langer1992, Bray2002, Cates2018}. These models are based on a free-energy functional expressed in terms of an order parameter $\phi$. In model A, where $\phi$ is not conserved, the dynamics is captured by a time-dependent Ginzburg-Landau equation, yielding $z=2$. In model B, where $\phi$  is conserved, the system evolves according to the Cahn-Hilliard (CH) equation, leading to $z=3$ (Lifshitz-Slyozov growth). However, real systems often exhibit deviations from these values due to various effects, such as the presence of topological defects in the BKT transition, or hydrodynamic flows  in classical binary liquids  \cite{Bray2002, Onuki2004}. Specifically, in binary liquids, the early-stage of the phase separation dynamics is typically described  by the  CH equation with $z=3$, while hydrodynamic flows were predicted  to induce crossovers at long times, first toward  a  viscous regime characterized by $z=1$ \cite{Siggia1979}, and eventually to an inertial regime with $z=3/2$ \cite{Furukawa1985}.

In the context of quantum physics, recent advances in the study of out-of-equilibrium ultracold gases have spurred systematic investigations into the dynamic scaling of correlations across phase transitions, both at the experimental \cite{Glidden2021, Galka2022, Bagnato2022, Abuzarli2022, Manovitz2025, Sunami2022, Gazo2023} and theoretical  \cite{Berges2008, Chantesana2019, Mikheev2019, Comaron2019, Marino2022, Cherroret2024, Zhu2023, Gliott2024, Scoquart2022, Kirkby2024} levels.
In particular, recently the non-equilibrium dynamics 
%question of coarsening has  emerged for 
of binary Bose \emph{superfluids} has sparked considerable interest, due to rich phenomelogy induced by the  intra- and inter-component repulsive interactions of the two species, of respective strengths $g$ and $g_{12}$ \cite{Schmied2019, Kudo2013, Hofmann2014, Singh2023, Hall1998, Papp2008, De2014, Qu2016, Cominotti2022, Prufer2018, Bresolin2023, Piekarski2025}. %. These systems are typically realized using ultracold gases with two spin components \cite{Hall1998, Papp2008, De2014, Cominotti2022, Prufer2018, Piekarski2025}, characterized by intra- and inter-component repulsive interactions of respective strengths $g$ and $g_{12}$. 
At zero temperature, binary Bose gases undergo a quantum phase transition at $g=g_{12}$: for $g>g_{12}$, the two components coexist (miscible phase), while for $g<g_{12}$, they separate (immiscible phase) \cite{Timmermans1998, Pu1998}. When quenched into the immiscible phase, numerical simulations have shown that binary Bose superfluids exhibit coarsening with a domain size growing as $L(t) \sim t^{2/3}$ \cite{Kudo2013, Hofmann2014, Singh2023}.
Physically, the observed exponent $z=3/2$ has been proposed to correspond to the long-time  regime of classical binary fluids \cite{Kudo2013, Hofmann2014, Singh2023}, where inertial terms in the Navier-Stokes equation become relevant in the absence of viscosity. This interpretation, however, implies that the coarsening process is driven by complex hydrodynamic flows in the superfluid mixture, ruling out a simple effective theory of the CH type. 
At the same time, numerical simulations of superfluid mixtures have revealed striking similarities with properties of the CH equation, such as a Porod's law for the structure factor \cite{Hofmann2014, Singh2023}. This raises a fundamental question: is superfluid coarsening truly dictated by hydrodynamic flows? 
In this paper, we show that the answer is no, the key reason being that superfluid mixtures cannot be simply viewed as classical fluid mixtures with zero viscosity. Instead, we find that the main mechanism governing superfluid coarsening is the competition between interactions and  \emph{quantum pressure}, a central ingredient of superfluids with no classical counterpart. We show that this competition is captured by an effective equation of motion (EOM) for the order parameter---the relative density difference $\phi$ between the two species, which we derive from the microscopic Hamiltonian describing the superfluid mixture. This equation bears some resemblance to the classical model B of statistical physics, but unlike the associated CH equation, which is of dissipative nature and of first-order in time, leading to $L(t) \sim t^{1/3}$, the effective EOM for binary superfluids is a conservative second-order differential equation and predicts $L(t) \sim t^{2/3}$. Our model also elucidates other aspects of superfluid coarsening, such as the Porod's law identified in previous works, and domain interfacial properties. Interestingly, hydrodynamic flows in the superfluid do not seem to play a dominant role, at least in a regime of weak segregation.

The paper is organized as follows. In Sec. \ref{Sec:theory}, we present a derivation of the effective EOM for binary Bose superfluids, starting from the microscopic Hamiltonian of the Bose mixture. We also discuss the corresponding effective Hamiltonian, for which energetically favorable configurations correspond to flat domains of the order parameter. Finally, we demonstrate that the EOM accurately captures the coarsening process in the immiscible phase.
As applications of the theory, we derive in Sec. \ref{Sec:applications} several key properties of superfluid coarsening: the growth law $L(t) \sim t^{2/3}$ and the associated domain-interface tension, Porod's law for the structure factor, and the shape of interfaces between domains. Comparisons between \emph{ab initio} simulations of Bose mixtures and the predictions of the effective theory, along with its range of validity, are discussed in Sec. \ref{Sec:validity}. Finally, Sec. \ref{Sec:conclusion} summarizes our findings and presents our conclusions.

\section{Effective theory of superfluid coarsening}
\label{Sec:theory}

\subsection{Hydrodynamics of Bose mixtures}

Consider a Bose gas made of two different species described by classical fields $\psi_i$, $i=1,2$. The Lagrangian density of the mixture reads (here and in the following, we set $\hbar=1)$
\begin{equation}
\mathcal{L}\!=\!i\psi_i^*\partial_t\psi_i-\frac{1}{2m}|\nabla\psi_i|^2-\frac{g}{2}|\psi_i|^4
-g_{12}|\psi_i|^2|\psi_j|^2,
\end{equation}
where $g$ denotes the intra-species interaction strength, which for simplicity we assume to be the same for the two species, and $g_{12}$ the inter-species interaction strength. By introducing the density-phase representation $\psi_i=\sqrt{\rho_j}e^{i\theta_i}$, we can rewrite this Lagrangian in terms of the hydrodynamic variables $\rho_i$ (density) and $\theta_i$ (phase) as:
\begin{equation}
\label{eq:L_hydro}
\mathcal{L}\!=\!\theta_i\partial_t\rho_i-\mathcal{H},
\end{equation}
with a Hamiltonian density
\begin{equation}
\label{eq:H_hydro}
\mathcal{H}\!=\!\frac{1}{2m}\Big[(\nabla\!\sqrt{\rho_i})^2\!+\!\rho_i(\nabla\theta_i)^2\Big]
\!+\!\frac{g}{2}\rho_i^2+g_{12}\rho_i\rho_j.
\end{equation}
This formulation leads to the well-known continuity and Euler equations of motion
\begin{align}
\label{EOM01}
&\partial_t\rho_i+\nabla(\rho_i\boldsymbol{v}_i)=0\\
\label{EOM02}
&\partial_t \boldsymbol{v}_i+\frac{1}{2}\nabla\boldsymbol{v}_i^2=-\frac{1}{m}\nabla\mu_i,
\end{align}
where we have introduced the velocities $\boldsymbol{v}_i=\nabla\theta_i/m$ of each component, as well as the chemical potentials
\begin{align}
\label{eq:mudef}
\mu_i=g\rho_i+g_{12}\rho_j-\frac{1}{2m}\frac{\nabla^2\sqrt{\rho_i}}{\sqrt{\rho_i}}.
\end{align}
The third, kinetic term on the right-hand side of this equation is known as the quantum pressure, and it will play a central role in the coarsening dynamics described in the subsequent sections.\newline

Let us now consider preparing, say at time $t=0$, a Bose mixture with balanced mean densities $\langle\rho_1 \rangle=\langle\rho_2 \rangle=\rho_0/2$ and zero mean velocities, $\langle\boldsymbol{v}_1\rangle=\langle\boldsymbol{v}_2\rangle=0$. We are interested in the time evolution of fluctuations around these averages. To describe these,  it is convenient to introduce two new variables, the normalized fluctuations of the total  density and density imbalance:
\begin{align}
\eta\equiv\frac{\rho_1+\rho_2-\rho_0}{\rho_0},\ \ \ 
\phi\equiv\frac{\rho_2-\rho_1}{\rho_0}.
\end{align}
We also define the total and relative phases, $\theta=\theta_1+\theta_2$ and $\delta\theta=\theta_2-\theta_1$, the corresponding velocities, $\boldsymbol{v}=\nabla\theta/m$ and $\delta\boldsymbol{v}=\nabla\delta\theta/m$, and finally the total and relative chemical potentials, $\mu=\mu_1+\mu_2$ and $\delta\mu=\mu_2-\mu_1$.
%In terms of these variables, the Lagrangian becomes:
%\begin{align}
%&\mathcal{L}\!=\!\frac{\hbar}{2}\Big(\theta\partial\rho+\Delta\theta\partial_t\Delta\rho\Big)+\frac{m\rho_0}{8}(\rho(\boldsymbol{v}^2+\Delta\boldsymbol{v}^2)+2\Delta\rho\boldsymbol{v}\Delta\boldsymbol{v}\Big)-\nonumber
%\\
%&\frac{\hbar^2}{4m\rho_0}\Big[(\nabla\sqrt{\rho\!-\!\Delta\rho})^2+(\nabla\sqrt{\rho\!+\!\Delta\rho})^2\Big]
%-\frac{1}{4}(g+g_{12})\rho^2
%-\frac{1}{4}(g-g_{12})\Delta\rho^2
%\end{align}
In terms of these variables, Eqs. (\ref{EOM01}) and (\ref{EOM02}) become
\begin{align}
\label{EOM1}
&\partial_t\phi+\frac{1}{2}\nabla\!\cdot\!\delta\boldsymbol{v}+\frac{1}{2}\nabla\!\cdot\!(\eta\delta\boldsymbol{v}+\phi\boldsymbol{v})=0\\
\label{EOM2}
&\partial_t\delta\boldsymbol{v}+\frac{1}{2}\nabla(\delta\boldsymbol{v}\cdot\boldsymbol{v})=-\frac{1}{m}\nabla\delta\mu\\
\label{EOM3}
&\partial_t\eta+\frac{1}{2}\nabla\!\cdot\!\boldsymbol{v}+\frac{1}{2}\nabla\!\cdot\!(\eta\boldsymbol{v}+\phi\delta\boldsymbol{v})=0\\
\label{EOM4}
&\partial_t\boldsymbol{v}+\frac{1}{4}\nabla(\boldsymbol{v}^2+\delta \boldsymbol{v}^2)=-\frac{1}{m}\nabla\mu,
\end{align}
where the total and relative chemical potentials $\mu$ and $\delta\mu$, deduced from Eq. (\ref{eq:mudef}), are functions of $\phi$ and $\eta$ only.

\subsection{Dynamical instability}

Let us now examine the behavior of fluctuations in the dynamical variables $\phi$, $\eta$, $\boldsymbol{v}$, and $\delta\boldsymbol{v}$ around their mean values $\langle\phi\rangle=0$, $\langle\eta\rangle=1$, and $\langle\boldsymbol{v}\rangle=\langle\delta\boldsymbol{v}\rangle=0$. We focus on the case where these fluctuations are initially small, and in this section we assume that they remain so over time (the validity of this assumption is discussed below). Linearizing Eqs. (\ref{EOM1}--\ref{EOM4}) then leads to two \emph{independent} sets of equations of motion for $(\phi, \delta\boldsymbol{v})$ and $(\eta, \boldsymbol{v})$, which describe the well-known spin and density modes of the Bose mixture \cite{Sinatra2000, Whitlock2003, Lellouch2013, Martone2021, Martone2023}. Eliminating the velocity variables further reduces these sets to two second-order time-dependent equations for $\phi$ and $\eta$,
\begin{align}
\label{eq:SD_waves}
\partial^2_t\phi&=\frac{\nabla^2}{2m}\Big[(g-g_{12})\rho_0-\frac{\nabla^2}{2m}\Big] \phi\\
\partial^2_t\eta&=\frac{\nabla^2}{2m}\Big[(g+g_{12})\rho_0-\frac{\nabla^2}{2m}\Big] \eta,
\end{align}
which typically describe periodic oscillations of the total density and density imbalance in space and time. In Fourier space, these equations read $(\partial^2_t+\Omega_{\bq,s}^2)\phi(\bq,t)=0$ and $(\partial^2_t+\Omega_{\bq,d}^2)\eta(\bq,t)=0$, where 
\begin{equation}
\Omega_{\bq,d/s}=\sqrt{\frac{\bq^2}{2m}\Big[\frac{\bq^2}{2m}+(g\pm g_{12})\rho_0\Big]}
\end{equation}
is the two-branch Bogoliubov dispersion for the density $(d,+)$ and spin $(s,-)$ modes. The latter reveals  that the spin mode---and consequently the linearization procedure---is stable only when $g_{12}<g$. This condition defines the miscible regime,  where both species coexist and occupy the entire available space. Conversely, when $g_{12}>g$, a dynamical instability emerges, leading to a divergence of the density imbalance $\phi(\br,t)$.  Mathematically, this divergence signals the breakdown of our initial assumption of weak fluctuations. Physically, the instablity arises due to a phenomenon of spinodal decomposition: the two species spontaneously separate and form domains. As a result, the density difference $\rho_2-\rho_1$ undergoes large fluctuations, while fluctuations in the total density $\rho_1+\rho_2$ remain typically small. 

\subsection{Effective equation of motion in the immiscible regime}

We now aim to construct a minimal model that captures the dynamics of the mixture in the immiscible regime $g_{12}>g$, starting from the exact hydrodynamic system (\ref{EOM1}--\ref{EOM4}). To this end, we build on the observation of the previous section: when $g_{12}\sim g$,  fluctuations of $\phi$ become large, preventing us from linearizing with respect to this variable. In contrast, fluctuations of $\eta$ remain \emph{a priori} small (this point will be examined in more detail in Sec. \ref{Sec:validity}). The behavior of velocity fluctuations, $\boldsymbol{v}$ and $\delta\boldsymbol{v}$, can be inferred from Eqs. (\ref{EOM1}) and (\ref{EOM3}), which  in the linear regime read  $\partial_t\phi\simeq -1/2\nabla\!\cdot\delta\boldsymbol{v}$ and $\partial_t\eta\simeq -1/2\nabla\!\cdot\boldsymbol{v}$. This indicates that at the onset of the dynamical instability, $\boldsymbol{v}$ keeps fluctuating weakly, whereas $\delta\boldsymbol{v}$ is tied to the growth of $\phi$. Based on these arguments, an effective EOM for the density imbalance in the immiscible regime can be derived by expanding Eqs. (\ref{EOM1}) and (\ref{EOM2}) with respect to $\eta$ and $\boldsymbol{v}$, but \emph{not} $\phi$ and $\delta\boldsymbol{v}$. This gives
\begin{equation}
\label{EOMf1}
\partial_t\phi\simeq-\frac{1}{2}\nabla\delta\boldsymbol{v}
\end{equation}
and  \begin{equation}
 \label{EOMf2}
\partial_t\delta\boldsymbol{v}\simeq-\frac{1}{m}\nabla\delta\mu,
\end{equation}
where $\delta\mu$ follows from Eq. (\ref{eq:mudef}):
\begin{equation}
\label{eq:dmudef}
\delta\mu\simeq (g-g_{12})\rho_0\phi-\frac{1}{2m}\left[\frac{\nabla^2\sqrt{1\!+\!\phi}}{\sqrt{1\!+\!\phi}}-\frac{\nabla^2\sqrt{1\!-\!\phi}}{\sqrt{1\!-\!\phi}}\right].
\end{equation}
Combining Eqs. (\ref{EOMf1}) and (\ref{EOMf2}) and simplifying the quantum-pressure term in Eq. (\ref{eq:dmudef}), we obtain the fundamental EOM describing the dynamics of the density imbalance in the immiscible regime:
\begin{equation}
\label{eq:effective}
\partial^2_t\phi\!=\!\frac{\nabla^2}{2m}\Big[(g\!-\!g_{12})\rho_0\phi+\frac{1}{2m}\frac{\nabla^2\text{Arccos}\phi}{\sqrt{1-\phi^2}}\Big].
\end{equation}
%\begin{equation}
%\label{eq:effective}
%\partial^2_t\phi\!=\!\frac{\nabla^2}{2m}\Big[(g\!-\!g_{12})\rho_0\phi-\frac{1}{2m}\Big[\frac{\nabla^2\phi}{1-\phi^2}\!+\!\frac{\phi(\nabla\phi)^2}{(1-\phi^2)^2}\Big]\Big].
%\end{equation}
Equation (\ref{eq:effective}) constitutes the central result of the paper. Compared to the Bogoliubov equation (\ref{eq:SD_waves}),  it still describes an evolution of the spin mode decoupled from the density mode, but now with a highly nonlinear dynamics. Moreover, while Bogoliubov theory is restricted to the miscible regime, Eq. (\ref{eq:effective}) captures the evolution in the immiscible regime $g_{12}>g$, where fluctuations of $\phi$ become significant. 
Equation (\ref{eq:effective}) is derived under the assumption of negligible fluctuations in the total density $\eta$, which constrains the density imbalance  $\phi=(\rho_2-\rho_1)/\rho_0\simeq(\rho_2-\rho_1)/(\rho_1+\rho_2)$  to lie within the interval $[-1, 1]$. Within this range, $\phi$ can, in principle, take any value. Importantly, Eq. (\ref{eq:effective}) neglects any phenomenon of hydrodynamic flow of the mixture, which is encoded in the nonlinear dynamics of the total velocity $\boldsymbol{v}$. Instead, Eq. (\ref{eq:effective}) describes a competition between interactions (first term on the right-hand side) and quantum pressure (second term), both independent of $\boldsymbol{v}$. More precisely, the negative interaction term $(g\!-\!g_{12})\rho_0\phi$ drives the dynamical instability, which is eventually counteracted by the positive quantum pressure term $\nabla^2\text{Arccos}\phi/\sqrt{1-\phi^2}$. As we will now show, this competition  underpins the coarsening dynamics, leading to the formation of sharp domain boundaries between the two superfluid components.

\subsection{Effective Hamiltonian and coarsening dynamics}

To gain deeper insight into the interplay between interactions and quantum pressure in the superfluid mixture, it is instructive to observe that the equations of motion (\ref{EOMf1}--\ref{eq:dmudef}), and consequently Eq. (\ref{eq:effective}), can also be derived from an effective Lagrangian formulation. This effective Lagrangian is obtained by expanding the microscopic Lagrangian, Eq. (\ref{eq:L_hydro}), to zeroth order in $\eta$ and $\boldsymbol{v}$:
\begin{equation}
\mathcal{L}_\text{eff}=\frac{\rho_0}{2}\delta\theta\,\partial_t\phi-\mathcal{H}_\text{eff},
\end{equation}
with an effective Hamiltonian density that only depends on the spin variables $\phi$ and $\delta\boldsymbol{v}$:
\begin{equation}
\label{eq:effectiveH}
\mathcal{H}_\text{eff}=\frac{\rho_0}{8}m\delta\boldsymbol{v}^2+\frac{\rho_0}{8m}\Big[\frac{(\nabla\phi)^2}{1-\phi^2}-\frac{\phi^2}{\xi_s^2}\Big],
\end{equation}
\begin{figure}[t!]
\includegraphics[scale=0.47]{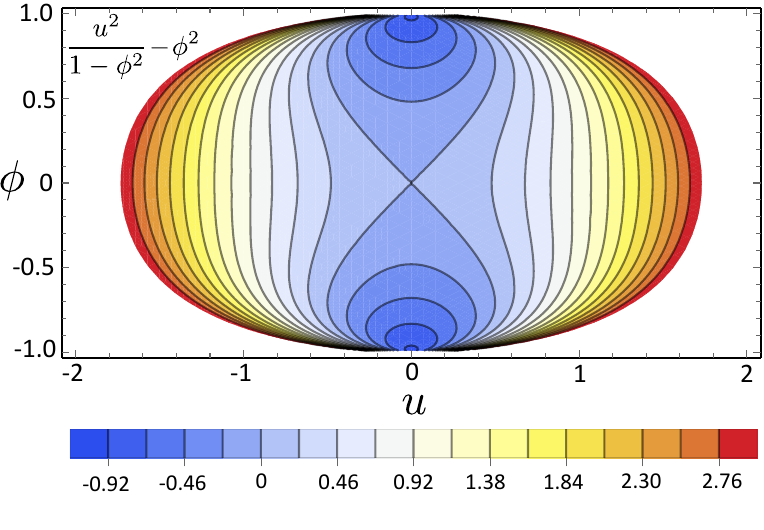}
\caption{\label{Fig:Potential}
Density plot of the interfacial potential $u^2/(1-\phi^2)-\phi^2/\xi_s^2$ appearing in the effective Hamiltonian density (\ref{eq:effectiveH}), here shown for $\xi_s=1$. The potential has a saddle point at $(u,\phi)=(0,0)$, and two global minima at $(u,\phi)=(0,\pm1)$ that correspond to the formation of domains.
}
\end{figure}
where we have introduced the `spin' healing length $\xi_s\equiv1/\sqrt{2m(g_{12}-g)\rho_0}$. Defining  $\Pi=\rho_0\delta\theta/2$, the conjugate momentum of $\phi$, it is easy to see that Hamilton's equations $\dot\phi=\delta H_\text{eff}/\delta\Pi$ and $\dot\Pi=-\delta H_\text{eff}/\delta\phi$ (with $H_\text{eff}=\int d^d\br \mathcal{H}_\text{eff}$) reduce to Eqs. (\ref{EOMf1}--\ref{eq:dmudef}).
Physically, the first term in the effective Hamiltonian (\ref{eq:effectiveH}) can be interpreted as the kinetic energy of the interface between two domains, while the second term represents  the interface's potential energy. This potential again highlights the competition between a positive quantum-pressure contribution, $(\nabla\phi)^2/(1-\phi^2)$, and a negative  interaction contribution, $-\phi^2/\xi_s^2$. The formation of domains resulting from this competition is reflected in the equilibrium points of the potential $u^2/(1-\phi^2)-\phi^2/\xi_s^2$, which is displayed in  Fig. \ref{Fig:Potential} in the $(u=\nabla\phi,\phi)$ plane: it exhibits two global minima at $(u,\phi)=(0,\pm1)$, corresponding precisely to the segregation of the species into flat domains with heights $\pm1$.
\begin{figure}[t!]
\includegraphics[scale=0.55]{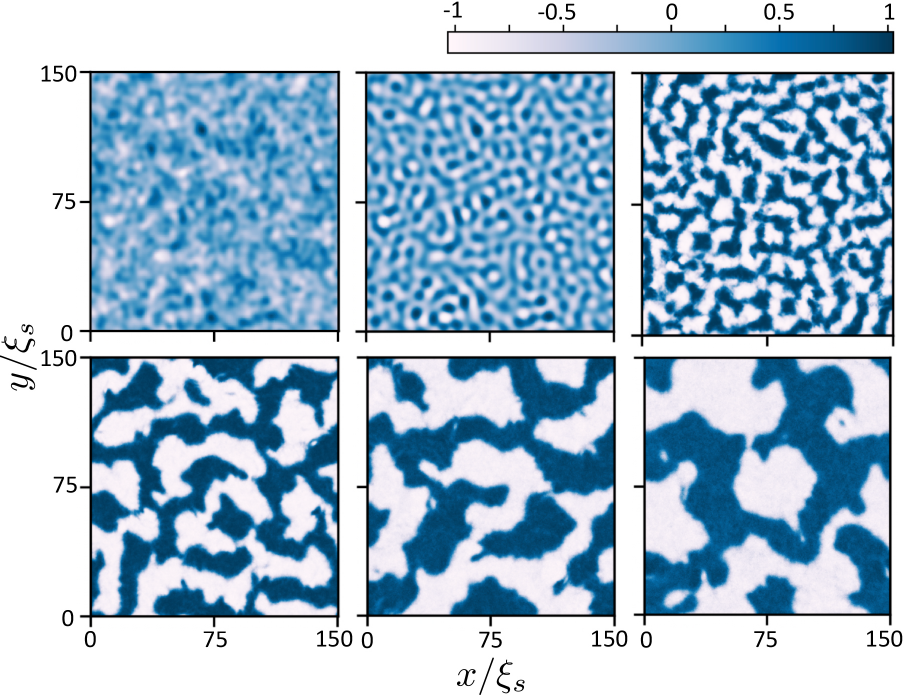}
\caption{\label{Fig:maps}
Density plots of the  density imbalance $\phi(x,y,t)$ at different times [from upper left to lower right: $t/t_\text{NL}=0, 10, 20, 60, 120, 180$]. The initial condition for $\phi(x,y,t)$ is   a uniformly distributed random field of zero mean and correlation function $\langle\phi(\br,0)\phi(\br',0)\rangle=\epsilon^2\exp[-(\br-\br')^2/4\sigma^2]$, where we here choose $\sigma=2\xi_s$ and $\epsilon=0.01$. Domains start to form around $t/t_\text{NL}\simeq 10$ and then grow in time.
%For the simulations, we discretize space in $1400$ grid points in each direction, and use a time step $dt=5.10^{-6}t_\text{NL}$.
}
\end{figure}

To  confirm that our theory does encapsulate superfluid coarsening, we have numerically investigated Eq. (\ref{eq:effective}). The latter can be conveniently written in dimensionless units as
\begin{equation}
\label{eq:effective_nodim}
\partial^2_{\tilde t}\phi\!=\!{\tilde\nabla^2}\Big[\phi+\frac{\tilde\nabla^2\text{Arccos}\phi}{\sqrt{1-\phi^2}}\Big],
\end{equation}
where $\tilde t\equiv t/t_\text{NL}$ and $\tilde x,\tilde y\equiv x/\xi_s, y/\xi_s$, with $t_\text{NL}\equiv 1/[(g_{12}-g)\rho_0$]. In Fig. \ref{Fig:maps}, we present spatial maps of the density imbalance $\phi(\br,t)$, obtained by numerically solving Eq. (\ref{eq:effective_nodim}) in two dimensions. The initial condition for $\phi$ is taken as a uniformly distributed real random field of zero mean and correlation function $\langle\phi(\br,0)\phi(\br',0)\rangle=\epsilon^2\exp[-(\br-\br')^2/4\sigma^2]$, with $\epsilon\ll1$. For details on the numerical procedure used to solve Eq. (\ref{eq:effective_nodim}), see Appendix \ref{Appendix:num}. We clearly observe the formation of domains between the two species, emerging on timescales of the order of a few $t_\text{NL}$ and subsequently growing. The coarsening of interfaces, which become increasingly sharper over time, is also evident. In the following sections, we will show how quantitative insights into this dynamics can be directly extracted from Eq. (\ref{eq:effective}) and the associated Hamiltonian (\ref{eq:effectiveH}).

\section{Applications of the theory}
\label{Sec:applications}

\subsection{Dynamical instability}

As a first investigation of Eq. (\ref{eq:effective}), we study the  growth of fluctuations of the density imbalance $\phi(\br,t)$ in the short-time limit where the dynamical instability occurs. In this regime, $\phi$ remains small so that the quantum-pressure term can be expanded for small $\phi$. The corresponding linearized equation coincides with the Bogoliubov equation (\ref{eq:SD_waves}), but where $g_{12}>g$. In Fourier space, it reads
\begin{equation}
\partial^2_t\phi(\bq,t)=\frac{1}{t_\text{NL}^2}(\bq^2\xi_s^2-\bq^4\xi_s^4)\phi(\bq,t),
\end{equation}
where $\phi(\bq,t)=\int d^d\br e^{i\bq\cdot\br}\phi(\br,t)$.
For an initial condition such that $\partial_t \phi(\bq,0)=0$ [which follows from Eq. (\ref{EOM1}) and our initial assumption that $\langle\boldsymbol{v}_1\rangle=\langle\boldsymbol{v}_2\rangle=0$], this yields the short-time solution:
\begin{align}
\phi(\bq,t)&=\phi(\bq,0)\Big[\cosh\Big(q\xi_s\sqrt{1\!-\!q^2\xi_s^2}\,\frac{t}{t_\text{NL}}\Big)\theta(1-q\xi_s)
\nonumber\\
&+\cos\Big(q\xi_s\sqrt{q^2\xi_s^2\!-\!1}\,\frac{t}{t_\text{NL}}\Big)\theta(q\xi_s-1)
\Big],
\end{align}
where $\theta$ is the Heaviside step function.
This relation can be used to infer the variance  $\langle\phi^2(\br,t)\rangle$. Using the same initial condition as in the simulation of Fig. \ref{Fig:maps}, % i.e.,  an initial real random field of correlation $\langle\phi(\br,0)\phi(\br',0)\rangle=\epsilon^2\exp[-(\br-\br')^2/4\sigma^2]$, 
we obtain:
\begin{align}
\label{eq:short-timeapprox}
&\langle\phi^2(\br,t)\rangle\!=\!\Big[
\int_0^{\xi_s^{-1}}\!\!\frac{qdq}{2\pi}e^{-q^2\sigma^2}\!\cosh^2\!\Big(q\xi_s\sqrt{1\!-\!q^2\xi_s^2}\,\frac{t}{t_\text{NL}}\Big)+\nonumber\\
&\int_{\xi_s^{-1}}^\infty\frac{qdq}{2\pi}e^{-q^2\sigma^2}
\cos^2\Big(q\xi_s\sqrt{q^2\xi_s^2\!-\!1}\,\frac{t}{t_\text{NL}}\Big)\Big] 4\pi\sigma^2\epsilon^2.
\end{align}
At times $t\gg t_\text{NL}$, the variance becomes dominated by the growing exponential in the second term of the right-hand side, which can be evaluated by a saddle-point approximation:
\begin{equation}
\label{eq:variance_SP}
\langle\phi^2(\br,t)\rangle\simeq\frac{\epsilon^2}{4}\sqrt{\frac{\pi}{2}}\left(\frac{\sigma}{\xi}\right)^2\frac{1}{\sqrt{t/t_\text{NL}}}\exp\left(\frac{t}{t_\text{NL}}\right).
\end{equation}
\begin{figure}[t!]
\includegraphics[scale=0.5]{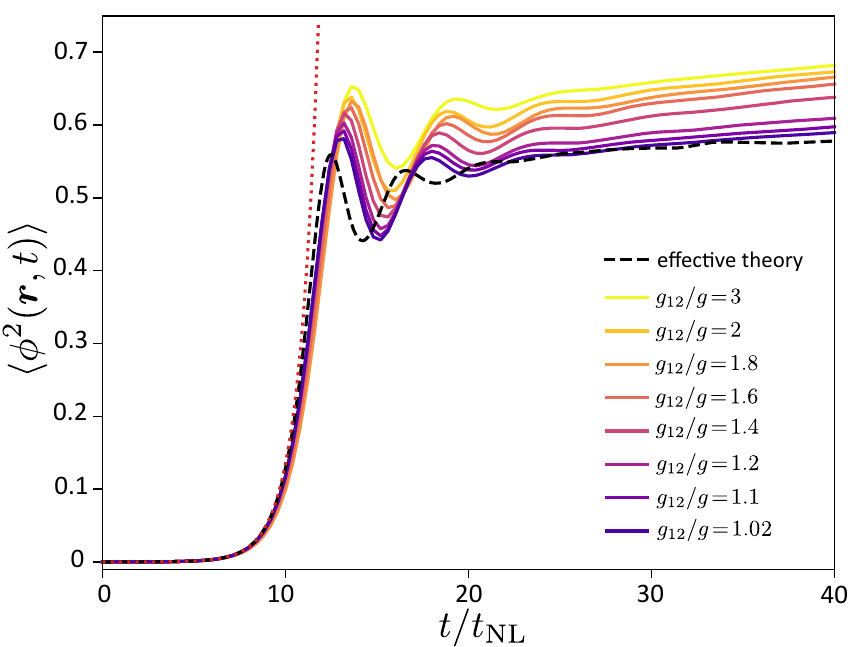}
\caption{\label{Fig:variance}
Variance of the density imbalance as a function of time $t/t_\text{NL}$. The dashed curve is the result of the effective EOM (\ref{eq:effective_nodim}), and the dotted curve the analytical prediction (\ref{eq:short-timeapprox}) for short times. Solid colored curves show results from \emph{ab initio} simulations for different values of the segregation parameter $g_{12}/g$, based on coupled nonlinear Shr\"odinger equations. Here $\sigma/\xi_s=2$.}
\end{figure}
In Fig. \ref{Fig:variance}, we show the variance of the density imbalance, numerically computed from Eq. (\ref{eq:effective_nodim}) (dashed curve). The curve initially exhibits a rapid growth, corresponding to the dynamical instability. This growth is well captured by the short-time prediction (\ref{eq:short-timeapprox}), shown as a dotted red curve. After a few $t_\text{NL}$, the instability is counterbalanced by the quantum pressure, leading to a much slower increase. The characteristic timescale $t_d$ at which this balance is established can be estimated by setting (\ref{eq:variance_SP}) to 1:
\begin{equation}
t_d\sim -t_\text{NL}\ln\Big(\frac{\epsilon\,\sigma}{\xi_s}\Big).
\end{equation}
Physically, $t_d$ corresponds to the time  when domains begin to form.

\subsection{Domain growth, Porod's law and interface tension}

\begin{figure}[t!]
\includegraphics[scale=0.55]{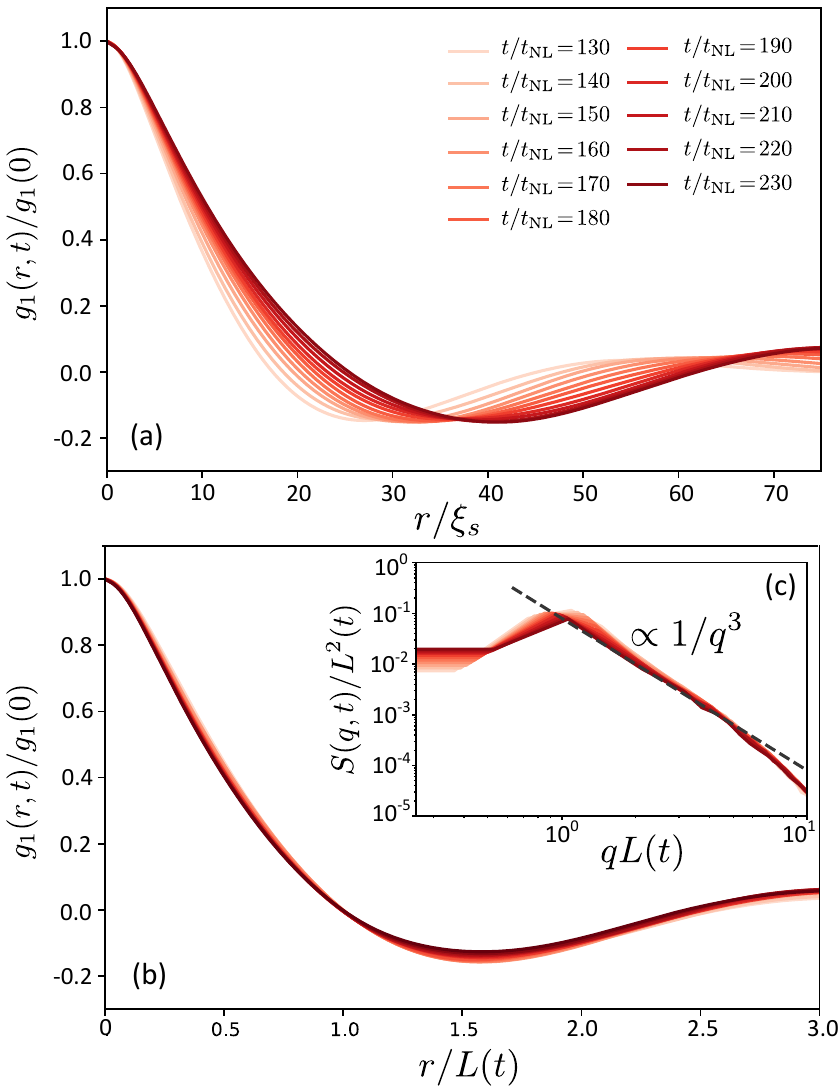}
\caption{\label{Fig:g1}
Correlation function $g(r,t)= \langle\phi(\br,t)\phi(0,t)\rangle$ at different times, numerically computed from from Eq. (\ref{eq:effective_nodim}). In (a), $g_1$ is shown as a function of $r/\xi_s$, and in (b) as a function of the rescaled position $r/L(t)$, where $L(t)$ is the first zero of $g$. The inset (c) shows the rescaled structure factor $S(q,t)/L^2(t)$ [with $S$ defined in Eq. (\ref{eq:Skdef})], as a function of the rescaled momentum $qL(t)$. The dashed line highlights the Porod's scaling $\sim 1/q^3$ expected for $1/L\ll q\ll1/\xi_s$.
Here $\sigma/\xi_s=2$, and the correlation function includes an angular average.
}
\end{figure}

\begin{figure}[t!]
\includegraphics[scale=0.55]{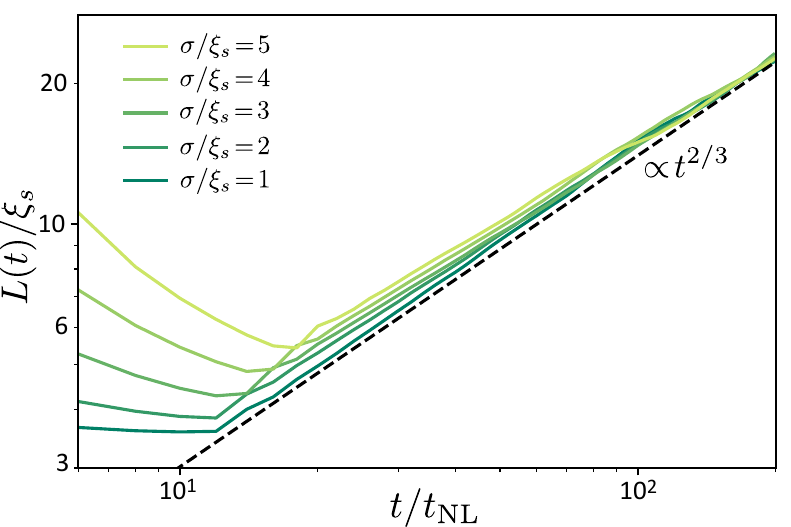}
\caption{\label{Fig:Loverxi}
Mean domain size $L(t)$ versus time, numerically computed  from the first zero of the correlation function $g(r,t)$, and for several values of $\sigma/\xi_s$ (with $\sigma$ the correlation length of the initial noise). At long times, all curves scale as $t^{2/3}$, with a prefactor independent of $\sigma/\xi_s$.
}
\end{figure}

We now focus on times $t\gg t_d$ where domains have formed and are growing. In this regime, the correlation function $g$ of the density imbalance is expected to follow a self similar dynamic scaling of the form \cite{Kudo2013, Hofmann2014, Singh2023}
\begin{equation}
\label{eq:dynamic_scaling}
g(r,t)\equiv \langle\phi(\br,t)\phi(0,t)\rangle=f\left[\frac{r}{L(t)}\right]
\end{equation}
where $r=|\br|$ and $L(t)$ is a characteristic length scale representing the mean domain size. To assess whether our effective theory predicts this scaling property and to determine $L(t)$, we have  computed $g(r,t)$ by numerically solving Eq. (\ref{eq:effective_nodim}). The results are shown in Fig. \ref{Fig:g1}(a) at different times.
By identifying $L(t)$ as the first zero of the $g$ function and replotting $g(r,t)$ as a function of $r/L(t)$, we obtain the curves in Fig. \ref{Fig:g1}(b). Remarkably, they all collapse onto a single profile, confirming the dynamic scaling law (\ref{eq:dynamic_scaling}). The resulting size $L(t)$  is shown in Fig. \ref{Fig:Loverxi} as a function of time, and is found to precisely follow the expected law $L(t)\propto t^{1/z}$ with $z={3/2}$.

As we now show, the value $z={2/3}$ naturally emerges from our effective theory and the associated EOM (\ref{eq:effective_nodim}). To demonstrate that $L(t)\propto t^{2/3}$, we present two distinct arguments.  The first, originally proposed in \cite{Huse1986} in the context of the Cahn-Hilliard equation, relies on dimensional analysis. Equation (\ref{eq:effective_nodim}) reads  $\partial^2_t\phi=(1/2m)\nabla^2\delta\mu$, where $\delta\mu$ represents the energy of an interface. The pressure difference $\rho_0\delta\mu$ sustained across the interface between two domains occupied by the species 1 and 2 is expected follow a Young-Laplace law:
\begin{equation}
\rho_0\delta\mu\sim \frac{\gamma}{L(t)}
\end{equation}
where $\gamma$ is a superfluid interfacial tension. By performing a dimensional analysis of the EOM, we infer $1/t^2\sim \gamma/(2mL^3\rho_0)$, which leads to
\begin{equation}
\label{eq:Lt_scaling}
L(t)\sim\Big(\frac{\gamma}{\rho_0 m}\Big)^{1/3}t^{2/3}.
\end{equation}
This provides a theoretical foundation for the scaling behavior previously observed in numerical studies \cite{Kudo2013, Hofmann2014, Singh2023}. Equation (\ref{eq:Lt_scaling}) can be viewed as an extension of the Lifshitz-Slyozov law  $L(t)\sim t^{1/3}$---predicted by the CH equation for classical fluids---to binary superfluids. 

A second, more rigorous argument for the scaling law $L(t)\propto t^{2/3}$ can be derived from the effective Hamiltonian density (\ref{eq:effectiveH}), which we rewrite as
\begin{equation}
\label{eq:effectiveH2}
\mathcal{H}_\text{eff}=\frac{\rho_0}{8}m\delta\boldsymbol{v}^2+\frac{\rho_0}{8m}\Big[(\nabla\phi)^2+\frac{\phi^2(\nabla\phi)^2}{1-\phi^2}-\frac{\phi^2}{\xi_s^2}\Big].
\end{equation}
In the coarsening regime, all terms in this expression compete with each other and thus scale similarly with time. In particular, we expect the mean kinetic energy to scale in the same way as the first term in the potential energy:
\begin{equation}
\langle E_\text{kin}(t)\rangle\sim \rho_0m\langle\delta\boldsymbol{v}^2\rangle\sim \langle E_\text{pot}(t)\rangle\sim \frac{\rho_0}{m}\langle(\nabla\phi)^2\rangle.
\end{equation}
Since $\langle\delta\boldsymbol{v}^2\rangle$ represents the mean velocity of interfaces, we have $\smash{\langle E_\text{kin}(t)\rangle\sim \rho_0 m[\dot L(t)]^2}$. On the other hand, the potential energy can be expressed as:
\begin{equation}
\label{eq:Epot}
\langle E_\text{pot}(t)\rangle\sim \frac{\rho_0}{m}\int\frac{d^2\bq}{(2\pi)^2}\bq^2S(\bq,t)
\end{equation}
where
\begin{equation}
\label{eq:Skdef}
S(\bq,t)=\int d^2\br\langle\phi(\br,t)\phi(0,t)\rangle e^{i\bq\cdot\br}
\end{equation}
is the structure factor. To estimate it, we focus on spatial scales $\xi_s\ll r\ll L(t)$, where we expect the motion of interfaces to dominate the correlation. Adapting the argument of \cite{Bray2002}, we estimate the probability of an interface being present between $0$ and $r$ as $r/L$, so that $\langle\phi(\br,t)\phi(0,t)\rangle\simeq (-1)r/L+(+1)(1-r/L)$. It follows that
\begin{equation}
\label{eq:Porodslaw}
S(\bq,t)\sim \int_{\xi_s\ll r\ll L}
\!\!\!\!\!\!\!\!\!\!\!\!\!\!\!\!d^2\br e^{i\bq\cdot\br}\Big(1-\frac{2r}{L}\Big)
\sim\frac{1}{q^3L(t)},
\end{equation}
which for classical binary fluids is known as Porod's law. The structure factor numerically computed from Eq. (\ref{eq:effective_nodim}) is shown in Fig. \ref{Fig:g1}(c), and indeed satisfies this law in the intermediate momentum range $1/L\ll q\ll 1/\xi_s$. Inserting Eq. (\ref{eq:Porodslaw}) into Eq. (\ref{eq:Epot}), we infer $\langle E_\text{pot}(t)\rangle\sim \rho_0/[m L(t) \xi_s]$. Comparing with the previous estimation for the kinetic energy, we deduce:
\begin{equation}
    \frac{\rho_0}{m\xi_s L(t)}\sim \rho_0 m \Big[\frac{dL(t)}{dt}\Big]^2.
\end{equation}
At long time, this  leads again to the growth law (\ref{eq:Lt_scaling}), with an interface tension given by
\begin{equation}
\label{eq:gamma}
    \gamma\sim\rho_0\sqrt{\frac{(g_{12}-g)\rho_0}{m}}.
\end{equation}
This expression is consistent with the previous works \cite{Ao1998, Mazets2002, Schaeybroeck2008}, in which  the interface tension was derived in the context of Bose mixtures at equilibrium in the weak segregation limit (see Sec. \ref{Sec:validity}).
We have also numerically verified this expression within our non-equilibrium approach. 
Specifically, by introducing the dimensionless variables $\tilde t=t/t_\text{NL}$ and $\tilde x,\tilde y=x/\xi_s,y/\xi_s$, Eq. (\ref{eq:Lt_scaling}) with $\gamma$ given by Eq. (\ref{eq:gamma}) implies a domain size scaling as $\tilde L(t)\propto\tilde t^{2/3}$, with a proportionality factor independent of any physical parameter. This property is confirmed in Fig. \ref{Fig:Loverxi}, which shows $\tilde L(t)$ versus $\tilde t$  on a log-log scale for various values of the initial noise correlation length: at long times, all curves converge to a single one.

\subsection{Shape of domain interfaces}

\begin{figure}
\includegraphics[scale=0.55]{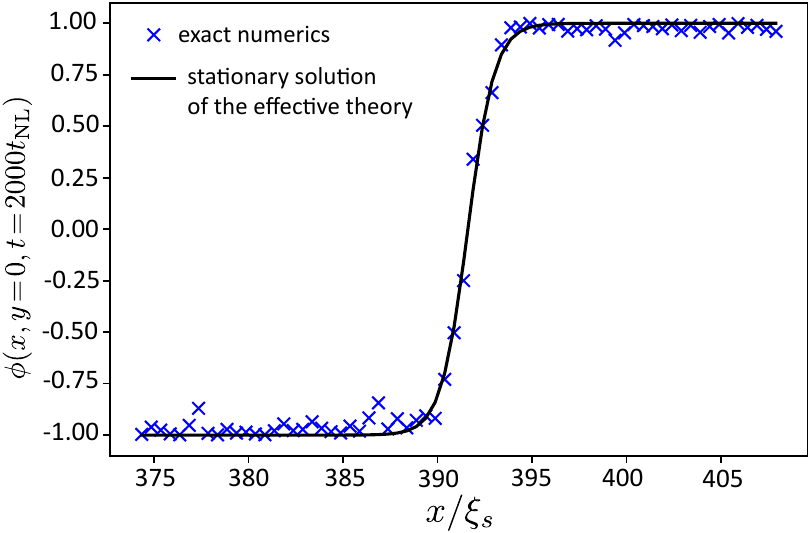}
\caption{\label{Fig:interface}
Transverse shape of an interface between two domains. Blue crosses are exact numerical results obtained by solving two-dimensional coupled Gross-Pitaevskii equations for $g_{12}/g = 1.1$ up to a time $t=2000 t_\text{NL}$, and the solid black curve is the prediction (\ref{Eq:interface}) from the effective theory, where the only fit parameter is the interface center $x_c\simeq 391.6\xi_s$.}
\end{figure}

As a last application of the  effective EOM (\ref{eq:effective}), we show that it can also be used to predict the shape of interfaces between domains. To do so, we note that at long times, the profile of interfaces becomes essentially time independent, so that it should match stationary (also known as `kink' \cite{Chaikin1995}) solutions of the EOM. These solutions can be found rather easily in one dimension, by imposing the right-hand side of Eq. (\ref{eq:effective}) to vanish. Expanding the second derivative of the Arccos function, this leads to \cite{Footnote}:
\begin{equation}
\label{eq:interfaceeq}
\frac{\phi}{\xi_s^2}+\frac{\partial^2_x\phi}{1-\phi^2}+\frac{\phi(\partial_x\phi)^2}{(1-\phi^2)^2}
=0,
\end{equation}
which can be rewritten as
\begin{equation}
    \frac{\partial}{\partial\phi}\left[\frac{(\partial_x\phi)^2}{1-\phi^2}+\frac{\phi^2}{\xi_s^2}\right]=0
\end{equation}
This relation is readily integrated as
\begin{equation}
    \frac{(\partial_x\phi)^2}{1-\phi^2}=\frac{\phi_0^2-\phi^2}{\xi_s^2},
\end{equation} 
with $\phi_0$ some integration constant. We are looking for a solution separating two domains that are flat far from the interface. We therefore require that $\partial_x\phi=0$ for $x\to\pm\infty$, which imposes $\phi_0^2=1$. We infer that $\partial_x \phi(x)=\pm (1-\phi^2)/\xi_s$, for which the solution is 
\begin{equation}
\label{Eq:interface}
\phi(x)=\pm\tanh\left(\frac{x-x_c}{\sqrt{2}\xi_s}\right).
\end{equation}
This relation describes a one-dimensional interface between two domains, centered around $x=x_c$, with $\phi(-\infty)=-1$ and $\phi(\infty)=+1$ (`kink' solution) or with $\phi(-\infty)=1$ and $\phi(\infty)=-1$ (`anti-kink' solution). In two dimensions, exact stationary solutions of Eq. (\ref{eq:effective}) are much more difficult to find, but since interfaces between domains are locally flat we expect their transverse cuts to be in very good approximation still described by Eq. (\ref{Eq:interface}). This is confirmed by Fig. \ref{Fig:interface}, which compares Eq. (\ref{Eq:interface}) to exact numerical results obtained from \emph{ab initio} simulations of coupled Gross-Pitaevskii equations \cite{Aladjidi2024} (see Appendix \ref{AppendixB} for details about these simulations).

\section{Role of hydrodynamic flows}
\label{Sec:validity}

The EOM (\ref{eq:effective}) has been derived under the assumption that spin and density degrees of freedom remain decoupled in the coarsening regime. Within this approximation, the nonlinear dynamics of the total velocity $\boldsymbol{v}$, which governs the center-of-mass hydrodynamic motion of the Bose mixture, does not influence the coarsening mechanism. However, it is important to determine the condition under which such hydrodynamic motion begins to play a role. 

In the previous works \cite{Qu2016, Congy2016}, it was suggested that a decoupling between spin and density degrees of freedom in the immiscible regime $g_{12}>g$ holds when $g_{12}/g$ is close to 1, i.e., near the transition. To verify whether this criterion also applies to the coarsening dynamics, we compare in Fig. \ref{Fig:variance} the variance of the order parameter computed from the effective theory with \emph{ab initio} simulations (see Appendix \ref{AppendixB}). The results confirm that the agreement with the effective theory is excellent when the ratio $g_{12}/g$ approaches 1, while deviations appear for $g_{12}/g>1$. To better understand the origin of these deviations, let us suppose that the dynamics of $\phi$ remains, to a first approximation, governed by Eq. (\ref{EOMf1}), i.e., $\partial_t\phi\sim \nabla\!\cdot\delta\boldsymbol{v}$,
and examine the consequence of a coupling between the dynamics of $\phi$ and that of the total density $\eta$. In Eq. (\ref{EOM3}), this coupling appears through the term $1/2\nabla\cdot(\phi\delta\boldsymbol{v})$. 
As long as the coupling remains weak, the total density is still governed by Eq. (\ref{eq:SD_waves}), ensuring that $\eta\ll1$. However, 
when the coupling becomes strong, the evolution of $\eta$ becomes tied to the fluctuations of $\phi$, causing the various terms in Eq. (\ref{EOM3}) to scale similarly in time. In particular:
\begin{equation}
\label{eq:estimate}
\partial_t\eta \sim 
\nabla\!\cdot\!\phi\,\delta\boldsymbol{v}\sim \phi\,\partial_t\phi\sim
\partial_t\phi^2.
\end{equation}
Since $\phi^2\sim 1$ is nearly constant in the coarsening regime at long time (see, e.g., Fig. \ref{Fig:variance}), the same is true for $\eta$, and we denote its long-time value by $\eta_\infty$. We now extrapolate  to short times, where $\partial_t \eta\sim (g+g_{12})\rho_0 \eta\sim (g+g_{12})\rho_0 \eta_\infty$ and $\partial_t \phi\sim (g_{12}-g)\rho_0 \phi\sim (g_{12}-g)\rho_0$ [see, e.g., Eq. (\ref{eq:variance_SP})]. Inserting these estimations in Eq. (\ref{eq:estimate}), we infer: 
\begin{equation}
\label{eq:etalong}
\eta_\infty\sim\frac{g_{12}-g}{g_{12}+g}=\frac{g_{12}/g-1}{g_{12}/g+1}.
\end{equation}
We conclude that $\eta_\infty$, and thus the coupling between spin and density degrees of freedom, remains small as long as $g_{12}/g$ is close to unity. This limit is known as the ``weak segregation'' regime, in which domain interfaces exhibit significant overlap between the two species \cite{Ao1998}. %This is the condition of validity of Eq. (\ref{eq:effective}).
\begin{figure}
\includegraphics[scale=0.62]{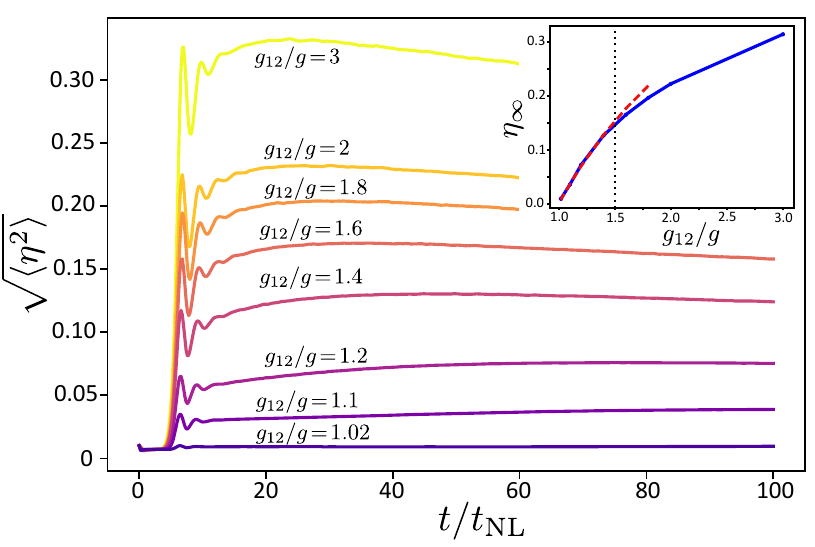}
\caption{\label{Fig:densityf}
Standard deviation of the total density $\eta$ as a function of time, computed from exact simulations of coupled Gross-Pitaevskii equations for different values of $g_{12}/g$. As $g_{12}/g$ increases, $\sqrt{\langle\eta^2\rangle}$ increases as well, confirming that spin and density degrees of freedom no longer decouple. The inset shows a long-time estimate $\eta_\infty$ of $\sqrt{\langle\eta^2\rangle}$ as a function of $g_{12}/g$. The dashed curve is a fit to Eq. (\ref{eq:etalong}), which works until $g_{12}/g\simeq 1.5$ (indicated by the vertical dotted line).
}
\end{figure} 
To validate these arguments, we have numerically computed the standard deviation $\sqrt{\langle\eta^2(\br,t)\rangle}$ as a function of time from \emph{ab initio} simulations, see Fig. \ref{Fig:densityf}. At long time, $\sqrt{\langle\eta^2\rangle}$ is indeed nearly constant, and its value, $\eta_\infty$, decreases as $g_{12}/g$ is approaches 1, indicating a weak coupling to $\phi$. The inset shows an estimate of the long-time value $\eta_\infty$, obtained by averaging  $\sqrt{\langle\eta^2\rangle}$ over times $t/t_\text{NL} > 10$, as a function of $g_{12}/g$. The observed behavior confirms the scaling predicted by (\ref{eq:etalong}) for small values of this ratio.

While a detailed study of the strong segregation regime falls beyond the scope of this paper, we have found that the coarsening dynamics is qualitatively similar when $g_{12}/g$ significantly differs from unity. In particular, the growth law $L(t)\sim t^{2/3}$ seems robust (slight changes in the exponent $2/3$ have been observed in \cite{Singh2023}, but it is presently not clear to us if they are representative or not), as well as the overall time evolution of the variance $\langle\phi^2(\br,t)\rangle$ (see Fig. \ref{Fig:variance}). This suggests a certain universality of the coarsening dynamics at large scales. On the other hand,  parameters  that are sensitive to the physics at small distance are 
typically renormalized for a strong segregation. This is the case, for instance, for the interface tension $\gamma$ and the interface size \cite{Ao1998}.

\section{Conclusion}
\label{Sec:conclusion}

In this paper, we developed an effective theory describing the coarsening dynamics of binary Bose gases, through the equation of motion (\ref{eq:effective}) or, equivalently, the effective Hamiltonian (\ref{eq:effectiveH}). This formulation is quantitatively accurate in the weak segregation regime and reveals that the fundamental mechanism driving superfluid coarsening is the competition between interactions and quantum pressure. It also offers an intuitive understanding of several key features of the coarsening process, such as domain growth and Porod’s law.

At the numerical level, a major challenge in solving the EOM (\ref{eq:effective}) arises from the highly nonlinear form of the quantum pressure, which becomes singular at the attractive points $\phi = \pm1$. Interestingly, however, a family of EOMs exhibiting dynamics similar to Eq. (\ref{eq:effective}), but free from such singularities, can be constructed by replacing the singular potential $u^2/(1 - \phi^2) - \phi^2/\xi_s^2$ (with $u \equiv \nabla\phi$) in the effective Hamiltonian (\ref{eq:effectiveH}) with an alternative potential that retains minima at the same points $u = 0$, $\phi = \pm1$, yet remains nonsingular. One such example is the potential $u^2\phi^2 - (\phi^2/\xi_s^2)(1 - \phi^2 + \phi^4/3)$, which fulfills these criteria. It leads to the alternative EOM
\begin{equation}
    \partial^2_t\phi\!=\!\frac{\nabla^2}{(2m)^2}\Big[
    \phi\nabla(\phi\nabla\phi)+
    \frac{1}{\xi_s^2}(\phi-2\phi^3+\phi^5)
    \Big],
\end{equation}
which should capture as well domain formation in a superfluid mixture, albeit probably less quantitatively than Eq. (\ref{eq:effective}). Such an approach is conceptually similar to strategies employed in classical binary fluids, where the singular logarithmic potential derived from thermodynamic considerations is often replaced by a quartic approximation, which eventually leads to the Cahn Hilliard equation \cite{Cherfils2011}.

The effective theory presented in this paper can be naturally extended to investigate several aspects of Bose superfluid coarsening that remain poorly understood. One such aspect concerns the nature of the dynamics in one dimension. In the case of the one-dimensional Cahn-Hilliard equation, it is well known that domain growth follows a much slower, logarithmic behavior \cite{Kawakatsu1985}, and we anticipate that a similar mechanism should arise in the superfluid counterpart (\ref{eq:effective}). Another intriguing direction is the case of unbalanced binary superfluids, where Ostwald ripening is expected to occur \cite{Fujimoto2020}, potentially leading to a modified domain growth law. Finally, the impact of quantum and thermal fluctuations on the superfluid coarsening dynamics remains, to our knowledge, largely unexplored. While thermal fluctuations have recently been shown to quantitatively affect growth laws \cite{Singh2023}, no analytical description of this process has yet been developed. Within our framework, we expect that incorporating an additional stochastic noise term into the EOM (\ref{eq:effective}) may provide a route to capture such effects.

\section{Acknowledgments}

Helpful discussions with Quentin Glorieux are gratefully acknowledged. This work has benefited from the financial support of Agence Nationale de la Recherche (ANR), France, under the Grant  No.~ANR-24-CE30-6695 FUSIoN. \newline

\appendix

\section{Numerical resolution of the effective equation  of motion}
\label{Appendix:num}

To  numerically solve Eq. (\ref{eq:effective_nodim}), we employ a semi-implicit pseudo-spectral scheme, previously used in the context of the Cahn–Hilliard equation \cite{Zhu1999}. To this aim, we rewrite Eq. (\ref{eq:effective_nodim}) in Fourier space as
\begin{equation}
\label{EOP:Fourier}
    \partial^2_{\tilde{t}} \hat{\phi}(\tilde\bq,\Tilde{t}) = \tilde{\bq}^2 [-\hat{\phi}(\tilde\bq,\Tilde {t})  + \hat{\mathcal{N}}(\phi)],
\end{equation}
where $\tilde{\bq} \equiv \bq\,\xi_s$, $\tilde{t} \equiv t/t_{NL}$, and $\hat{\phi}(\tilde\bq,\tilde{t})$ denotes the two-dimensional Fourier transform of $\phi(\tilde\br,\tilde{t})$. The term $\hat{\mathcal{N}}(\phi)$ represents the Fourier transform of the nonlinear quantum pressure contribution, which we express as
\begin{equation}
    \mathcal{N}(\phi) = \frac{\mathcal{F}^{-1}[\tilde{\bq}^2 \mathcal{F}(\text{Arccos}\,{\phi}(\tilde\br,\Tilde{t}))]}{\sqrt{1 - \phi^2(\tilde\br,\tilde{t})}},
\end{equation}
where $\mathcal{F}^{-1}$ denotes the inverse Fourier transform.  The main numerical challenge in evaluating $\mathcal{N}(\phi)$ arises from the singular behavior of both the denominator $\sqrt{1 - \phi^2}$ and the numerator $\mathcal{F}^{-1}[\tilde{\bq}^2 \mathcal{F}(\text{Arccos}\,{\phi})]$ when $\phi$ approaches $\pm 1$. To regularize these singularities, we approximate $1/(1-\phi^2)$ by its Taylor expansion up to order 110, and replace $\text{Arccos}\,\phi$ with its Pad\'e approximant of order $(m,n)=(8,8)$.
We then apply a semi-implicit time integration scheme, where the linear term on the right-hand side of Eq. (\ref{EOP:Fourier}) is treated implicitly, while the nonlinear term  $\hat{\mathcal{N}}(\phi)$ is evaluated explicitly. Discretizing the second-order time derivative yields:
\begin{equation}
    \hat{\phi}_{n+1} = \frac{2\hat{\phi}_{n} - \hat{\phi}_{n-1}  + \Delta \tilde{t}^2\, \tilde{\bq}^2\, \hat{\mathcal{N}}(\phi_{n})}{1 - \Delta \tilde{t}^2\, \tilde{\bq}^2},
\end{equation}
where $\hat{\phi}_n\equiv \hat{\phi}(\tilde\bq,\Tilde{t}_n)$ and $\Delta {\tilde{t}} = \Tilde{t}_{n+1} - \Tilde{t}_n$ is the time step. For the initial step, we use the condition that the time derivative vanishes at $t=0$, implying $ \hat\phi_{-1}=\hat\phi_1$. The stability of the numerical scheme requires that $\Delta \tilde{t} \ll \Delta \tilde{x}^4$, where $\Delta \tilde{x}$ is the spatial grid spacing. Compared to the fully explicit Euler method, this semi-implicit approach offers significantly improved numerical stability.

All numerical simulations of Eq. (\ref{eq:effective_nodim}) presented in the paper use, as an initial condition, a  real random field $\phi(\tilde\br,\tilde{t})$ uniformly distributed, with zero mean and spatial correlation $\langle\phi(\tilde\br,0)\phi(\tilde\br',0)\rangle=\epsilon^2\exp[-(\tilde\br-\tilde\br')^2/4\tilde\sigma^2]$, where $\epsilon = 0.01$. The parameter $\tilde\sigma\equiv\sigma/\xi_s$ is adjustable.
The grid size is fixed to $x_{\text{max}}/\xi_s = y_{\text{max}}/\xi_s = 150$, while the number of grid points in each direction, $N_x = N_y = N$, is chosen to meet the precision requirements of each observable: $N=1024$ for Fig. \ref{Fig:g1}, $N=1400$ for Figs. \ref{Fig:maps} and \ref{Fig:Loverxi}, and $N=1800$ for Fig. \ref{Fig:variance}. The time-step $\Delta \tilde{t}$ is then adjusted to meet stability requirements : $\Delta \tilde{t} = 2 \times 10^{-5}$ for simulations with $N=1024$ and $\Delta \tilde{t} = 1 \times 10^{-5}$ for all others. Simulation results are averaged over $n$ independent realizations of the initial random field, where $n=8$ in Fig. \ref{Fig:variance}, $n=7$ in Fig. \ref{Fig:Loverxi} and $n=12$ in Fig. \ref{Fig:g1} (due to the spatial averaging process in the calculation of the observables $\langle \phi(\br,t)^2 \rangle$ and $g_1(r,t)$, it is not necessary to average over many initial configurations).

\section{Numerical resolution of the coupled nonlinear Schrödinger equations}
\label{AppendixB}

In Figs. \ref{Fig:variance}, \ref{Fig:interface}, and \ref{Fig:densityf}, we present numerical solutions of the full coupled nonlinear Schrödinger equations:
\begin{align}
\label{GP1}
i \hbar \partial_{t} \psi_1 &= -\frac{\hbar^2}{2m} \mathbf{\nabla}^2 \psi_1 + g\lvert \psi_1 \rvert^2 \psi_1 + g_{12} \vert \psi_2 \rvert^2 \psi_1,\\
\label{GP2}
i\hbar \partial_{t} \psi_2 &= -\frac{\hbar^2}{2m}  \mathbf{\nabla}^2 \psi_2 + g\lvert \psi_2 \rvert^2 \psi_2 + g_{12} \vert \psi_1 \rvert^2 \psi_2,
\end{align}
which conserve the total density: $\int d^2\br(\lvert \psi_1 \rvert^2+\lvert \psi_2 \rvert^2)=1$.
Equations (\ref{GP1}) and (\ref{GP2}) are solved using a spectral split-step method implemented in the  package \cite{Aladjidi2024} developed by Aladjidi et al. We fix the parameters $\hbar = m = 1$. The grid spacing $dx$ in both directions is arbitrarily chosen as $dx \simeq 0.0111$ for \ref{Fig:variance} and \ref{Fig:densityf} and $dx \simeq 0.0028$ in \ref{Fig:interface}. The grid size is then fixed to $x_{\text{max}} = y_{\text{max}} = N dx$ with $N = 1024$. The timestep in all simulations is $dt = 0.004 t_0$ with $t_0 = 2m dx^2/\hbar$. The two fields are initialized as : 
\begin{align}
    \psi_1(\br,0) &= \sqrt{\rho_0/2}, \\
    \psi_2(\br,0) &= \sqrt{\rho_0/2}\,[1 + \epsilon\, \eta(\br)]/\sqrt{1 + \epsilon^2},
\end{align}
where $\eta$ is a random noise such that $\langle\eta(\br)\eta(\br')\rangle=\exp[-(\br-\br')^2/4\sigma^2]$. In all simulations the correlation length is  fixed to $\sigma = 4 dx$ and the mean total density $\rho_0 \simeq 9.54 \times 10^{-7} (1/dx^2)$. For $\epsilon \ll 1$, the density imbalance $\phi \equiv (\lvert \psi_2 \rvert^2 - \lvert \psi_1 \rvert^2)/\rho_0$ has thus the same initial condition as for the effective theory simulation detailed in appendix \ref{Appendix:num} : $\langle\phi(\br,0)\phi(\br',0)\rangle = \epsilon^2\exp[-(\br-\br')^2/4\sigma^2]$, where $\epsilon = 0.01$ and $\sigma/\xi_s = 2$. The value of $(g_{12} - g)\rho_0 = 1/(2m\xi_s^2)$ is fixed to $1/(8 dx^2)$ since $\xi_s = 2 dx$ and $g\rho_0 = (g_{12} - g)\rho_0/(g_{12}/g - 1)$ is adjusted according to the different $g_{12}/g$ ratios in \ref{Fig:variance} and \ref{Fig:densityf}. The results in both Figs. \ref{Fig:variance} and \ref{Fig:densityf} are averaged over $8$ independent realizations of the initial random field.


\begin{thebibliography}{99}
\bibitem{Footnote}
Strictly speaking, requiring the right-hand side of Eq. (\ref{eq:effective}) to vanish only imposes that $\partial_x^2\delta\mu=0$. This implies that $\delta\mu=ax+b$, where $a$ and $b$ are integration constants. However, since we are considering a balanced mixture,  $\phi=0$ must be a solution, which in turn requires $a=b=0$ and leads to Eq. (\ref{eq:interfaceeq}).
\end{thebibliography}
\end{document}